\documentclass[11pt,preprint,a4paper]{aastex} 

\newcommand{\ms}{\mbox{m\,s$^{-1}$}}

\newcommand{\kms}{\mbox{km\,s$^{-1}$}}
\newcommand{\msun}{M$_{\odot}$}

\newcommand{\mjup}{M$_{\rm Jup}$}

\newcommand{\mearth}{M$_{\rm Earth}$}

\newcommand{\msini}{$m \sin i$}

\shortauthors{O'Toole {\it et~al.\/}}
\shorttitle{HD 16417}
\slugcomment{ApJ   Submitted:                 Accepted:             }
\received{}
\accepted{}
\begin{document}

\title{A Neptune-mass Planet Orbiting the Nearby G Dwarf HD 16417$~^{1}$}

\author{
Simon O'Toole\altaffilmark{2},
C. G. Tinney\altaffilmark{3},
R. Paul Butler\altaffilmark{4}, 
Hugh R. A. Jones\altaffilmark{5},
Jeremy Bailey\altaffilmark{3},
Brad D. Carter\altaffilmark{6},
Steven S. Vogt\altaffilmark{7},
Gregory Laughlin\altaffilmark{7},
Eugenio J. Rivera\altaffilmark{7}
}

\email{otoole@aao.gov.au}

\altaffiltext{1}{Based on observations obtained at 
the Anglo-Australian Telescope, Siding Spring, Australia.}

\altaffiltext{2}{Anglo-Australian Observatory, P.O. Box 296,
Epping, NSW 1710, Australia}

\altaffiltext{3}{Department of Astrophysics, School
of Physics, University of New South Wales, NSW 2052,
Australia}

\altaffiltext{4}{Department of Terrestrial Magnetism,
Carnegie Institution of Washington, 5241 Broad Branch Road NW,
Washington D.C. USA 20015-1305}

\altaffiltext{5}{Centre for Astrophysical Research,
University of Hertfordshire, Hatfield, AL10 9AB, UK}

\altaffiltext{6}{Faculty of Sciences, University of Southern
Queensland, Toowoomba, Queensland 4350, Australia}

\altaffiltext{7}{UCO/Lick Observatory, University of California at 
Santa Cruz, CA 95064.}

\begin{abstract} Precision Doppler measurements from an intensive
48 night ``Rocky Planet Search'' observing campaign on the 
Anglo-Australian Telescope (AAT) have
revealed the presence of a low-mass exoplanet orbiting the G1 dwarf
HD\,16417. Subsequent Doppler observations with the AAT, as well
as independent observations obtained by the Keck Planet Search, have
confirmed this initial detection and refine the orbital parameters
to period 17.24$\pm$0.01\,d, eccentricity
0.20$\pm$0.09, orbital semi-major axis 0.14$\pm$0.01\,AU and
minimum planet mass 22.1$\pm$2.0\,\mearth. HD\,16417 raises the number of
published exoplanets with minimum masses of less than 25\,\mearth\ to
eighteen. Interestingly, the distribution of
detected sub-25\mearth\ planets over the spectral types G, K and M is almost
uniform. The detection of HD\,16417b by an intensive observing
campaign clearly demonstrates the need for extended and contiguous observing
campaigns when aiming to detect low-amplitude Doppler planets in
short period orbits. Perhaps most critically it demonstrates that the
search for low-mass Doppler planets will eventually require these traditional 
``bright-time'' projects to extend throughout dark lunations.
\end{abstract}


\section{Introduction} \label{intro}

Pushing exoplanet detection thresholds down to lower and lower masses has
been a significant science driver in exoplanetary science in recent
years. Seventeen Doppler exoplanets have been published to date with minimum
(i.e. \msini) masses of less than 25\,\mearth\ -- 
GJ\,426b \citep{Butler04}; 
HD\,219826b \citep{Melo07}; 
HD\,69830b,c,d \citep{Lovis06};
HD\,190360c \citep{Vogt05};
Gl\,581b,c,d \citep{Udry07};
HD\,4308b \citep{Udry06}
HD\,160691d \citep{Santos04};
Gl\,674b \citep{Bonfils07};
55\,Cnc\,e \citep{McArthur04};
Gl\,876d \citep{rivera05};
and the recently announced HD\,40307b,c,d \citep{Mayor08}.
The majority of these have been found in short period orbits, with the 
only three (HD\,69830c,d and Gl\,581d) having 
orbital periods greater than 30\,d. Of these eighteeen low-mass
exoplanets, roughly equal numbers have been
found orbiting M-, K- and G-dwarfs (6, 6 and 5
respectively in each of these spectral types). 
A further three microlensing planets with masses in this range
have also been detected -- in each case orbiting stars of M-class or later 
\citep{Gould06,Beaulieu06,Bennett08} because these stars dominate the
field star population that microlensing surveys probe.

The roughly equal distribution with host spectral type for the
low-mass Doppler exoplanets hides several selection
effects. First that finding very low-mass planets orbiting G-dwarfs is {\em
much} harder than finding them orbiting M-dwarfs, since the lower mass of
an M-dwarf primary will (for a given mass planet of a given orbital
period) make the Doppler amplitude of an M-dwarf exoplanet at
least three times larger than a G-dwarf one. And second, that current
planet search target lists are dominated by G-dwarfs.

The detection of such low-mass exoplanets within the last 4-5 years, has
in large part been due to the dramatic improvements achieved in the
intrinsic, internal measurement  precisions of Doppler planet search
facilities. These have  improved to such an extent, that it is now
clear that noise sources {\em intrinsic} to the parent star themselves are
the limiting factor for very low-mass exoplanet detection. Characterization of
these noise sources (jitter, convective granulation and
asteroseismological p-mode oscillations) has become an important focus of
Doppler planet detection. A few obvious modifications to current
observing strategies have emerged -- (1) target low-mass stars; (2)
target chromospherically inactive and slowly rotating stars; (3)
target high-gravity stars (where p-mode oscillations are minimized) and
(4) extend the observations of stars over several p-mode fundamental
periods, so that asteroseismological noise is averaged over.

In this paper, we present first results from a major observing campaign
-- the Anglo-Australian Rocky Planet Search -- that focussed on the last
three of these observing strategies, in an effort to push to the lowest
possible detection limits achievable with the Anglo-Australian Planet
Search (AAPS)  Doppler system. The AAPS began operation in 1998
January, and is currently surveying 250 stars.  It has first discovered
thirty-one exoplanets with \msini\ ranging from 0.17 to 10\,\mjup\
\citep{AAPSI,AAPSIII,AAPSVII,AAPSXI,AAPSXIII,AAPSII,AAPSV,
AAPSIV,AAPSVI,AAPSVIII,AAPSXII,AAPSIX,AAPSX,AAPSXIV,AAPSXV}.

The Anglo-Australian Rocky Planet Search targets unevolved dwarfs with
low activity levels from our main AAPS program. The observing strategy is
to observe every target, on every night of a contiguous 48 night observing
run (modulo, of course, the vagaries of weather). 
This 48n observing run covered two bright lunations, and included
an entire dark lunation. Each observation extends over at
least 15 minutes in order to beat down p-mode oscillation noise to levels
well below 1\,\ms\ (O'Toole et al. 2008).  The full Rocky Planet
Search target  list includes 55 objects, of which 24 were targeted on
our first 48n campaign in 2007 Jan \& Feb.

In this letter we present results for the most compelling new
exoplanetary detection (\object[HD16417]{HD\,16417b}) to arise from this 
concentrated, campaign-mode observing run, together with subsequent AAPS
and Keck Planet Search observations.

\section{HD\,16417} \label{16417}

HD16417 (GJ 101.1, HIP 12186) lies at a distance of 25.5$\pm$0.4\,pc 
\citep{Perry97}, and has a spectral type of G1V
\citep{Houk82,Gray06}, an absolute magnitude of $M_{\rm V} =
3.74$ ($V=5.78$)  and colour $B-V=0.653$.  Hipparcos photometry finds it to be
photometrically stable  at the 7 milli-magnitude level over 212
observations over the course of the Hipparcos mission \citep{Perry97}.  

As a bright and nearby Sun-like star, HD\,16417 has been the subject of
multiple detailed  atmospheric and isochrone analyses -- the conclusions
reached by the most recent of these are summarized in Table
\ref{hd16417_atm}. The first point to notice is that all these analyses
agree that, while the gravity of HD\,16417 is not low enough for it to be
classified as a giant or sub-giant, it is somewhat lower than the $\log g
\approx 4.5$ one would expect from a main sequence early-G dwarf,
indicating that it has begun to evolve off the main sequence. Where ages
have been estimated, they indicate HD\,16417 to be somewhat older than
the Sun -- in  the range 4-8\,Gyr. The mass of HD\,16417 is estimated to
be somewhat larger than that for the Sun, with the most recent estimates
of \citet{ValentiFischer05} being 1.38 and
1.18\,\msun\ (based, respectively, on spectroscopic analysis and isochrone analysis)
and that of \citet{daSilva06} being 1.18\,\msun. In the analysis that follows we assume a mass
of 1.2\,\msun.

Metallicity estimates for HD\,16417 range from [Fe/H] of $-$0.01 to
$+$0.19, with an average value of [Fe/H]=$+$0.06. Perhaps most critically
for the purposes of this study, HD\,16417 is a slow rotator ($v \sin i =
2.1$\,\kms) and extremely inactive (log\,R$^\prime_{\rm HK}=-5.08$), making it
an ideal target for Doppler planet searching at very high precision. The
updated \ion{Ca}{2} jitter calibration of J.Wright (priv.comm.) for
HD16417 indicates a jitter of 2.2\,\ms. The somewhat lower gravity of
HD16417 than solar indicates that Doppler observations will by slightly
affected by Doppler noise due to p-mode oscillations; the
relations of \citet{otoole08a} indicate an rms noise equivalent  
of less than 0.6\,\ms, for
observations of more than 10 minutes.

\section{Observations}
\label{obs}

AAPS Doppler measurements are made with the UCLES echelle spectrograph
\citep{diego:90}.  An iodine absorption cell provides wavelength
calibration from 5000 to 6200\,\AA.  The spectrograph point-spread
function and wavelength calibration is derived from the iodine absorption
lines embedded on every pixel of the spectrum by the cell \citep{val:95,BuMaWi96}.

Observations of HD\,16417 began as part of the AAPS main program in 1998,
and over the following seven years it was observed regularly in
observations of 300-600s (depending on observing conditions), giving a
signal-to-noise ratio (SNR) of $\approx$200 per spectral resolution
element in the iodine region. These are the observations listed in Table
\ref{velhd16417} between JD=2450831.0428 - 245381.1880. In 2005 Jul,
HD\,16417 (together with a number of other bright AAPS targets) was
elevated within our observing program to high-SNR status, such that its
target SNR per epoch became 400 per spectral pixel. The result was that
the median internal uncertainties produced by our Doppler fitting process
dropped from 1.59\,\ms\ to 0.77\,\ms. This improvement gave us confidence that our
Rocky Planet Search strategy (concentrating on a small number of targets
observed as contiguously as possible over a long observing run) would
significantly lower our noise levels for the detection of low-mass
planets. 

Observations for our Rocky Planet Search program began on 2007 Jan 10 and
continued through 2007 Feb 26. HD\,16417 was able to be observed on 24 of
those nights. Since this 48 night run, it has been
observed on a further ten nights at the AAT, and on ten nights  on the
Keck~I telescope with HIRES \citep{Vogt94}. The Doppler velocities
derived  from all these observations are  listed in Table
\ref{velhd16417}.

\section{Analysis}

The root-mean-square (rms) scatter about the mean velocity of all data
taken before 2005 Jul was 4.9\,\ms. The rms about the mean for all
data taken since that date is 4.4\,\ms. This is slightly smaller than
that seen in the  earlier, lower SNR data, but is
still significantly larger than would be expected based on
the internal measurement uncertainties, and the noise from jitter and p-modes
in this star. However, in spite of being observed with this improved
precision at 16 epochs over the period 2005 July to 2006 November, no
convincing periodicity could be extracted from the resultant Doppler
velocities.

The 48 night run in 2007, however, provided a 
clear indication of periodicity at $\approx$ 17\,d. It was then prioritized for
intensive observation over the following 18 months, and subsequent
data has confirmed the detection first made in our large,
contiguous observing block.

\begin{figure}
\includegraphics[clip=true,width=8cm]{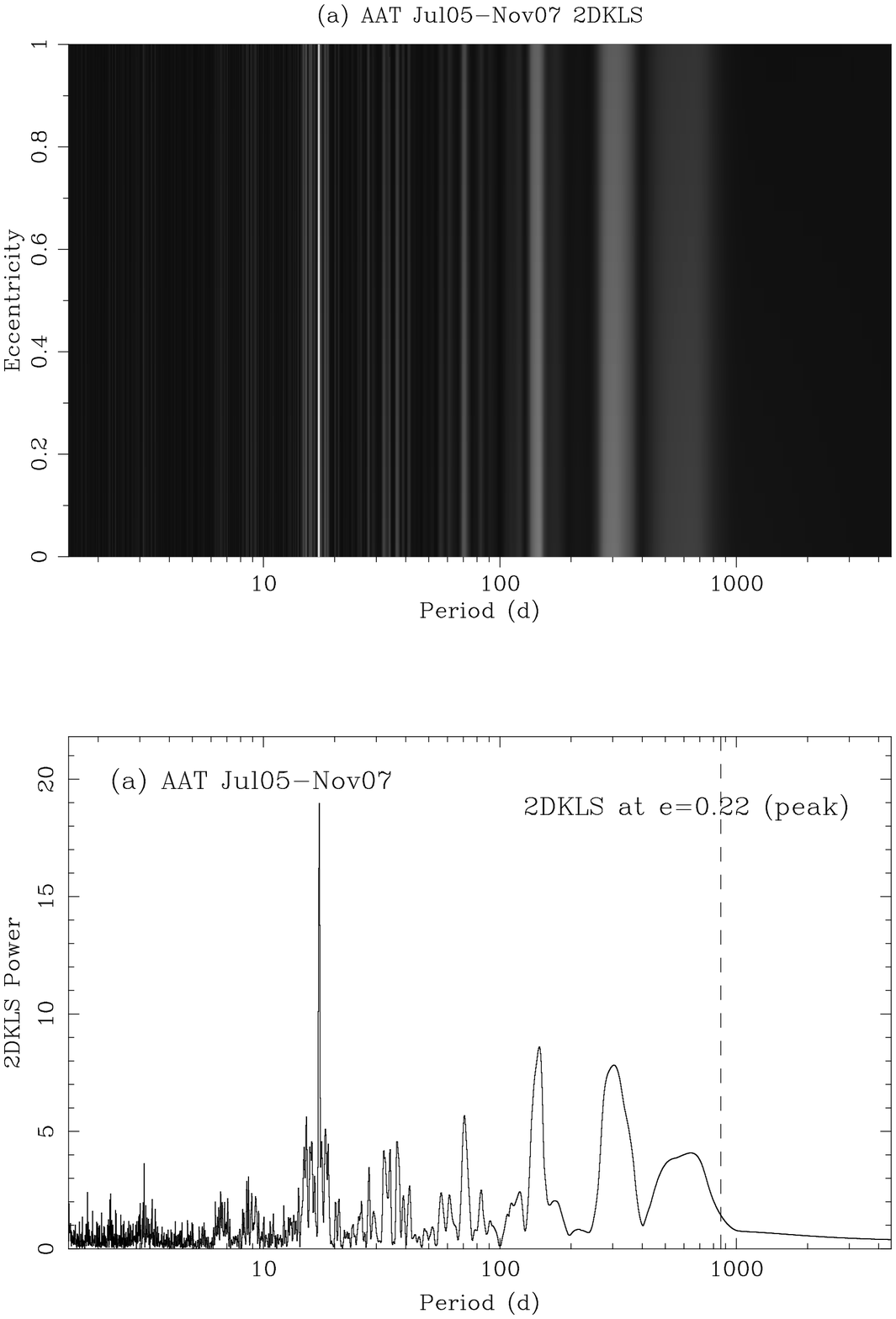}
\includegraphics[clip=true,width=8cm]{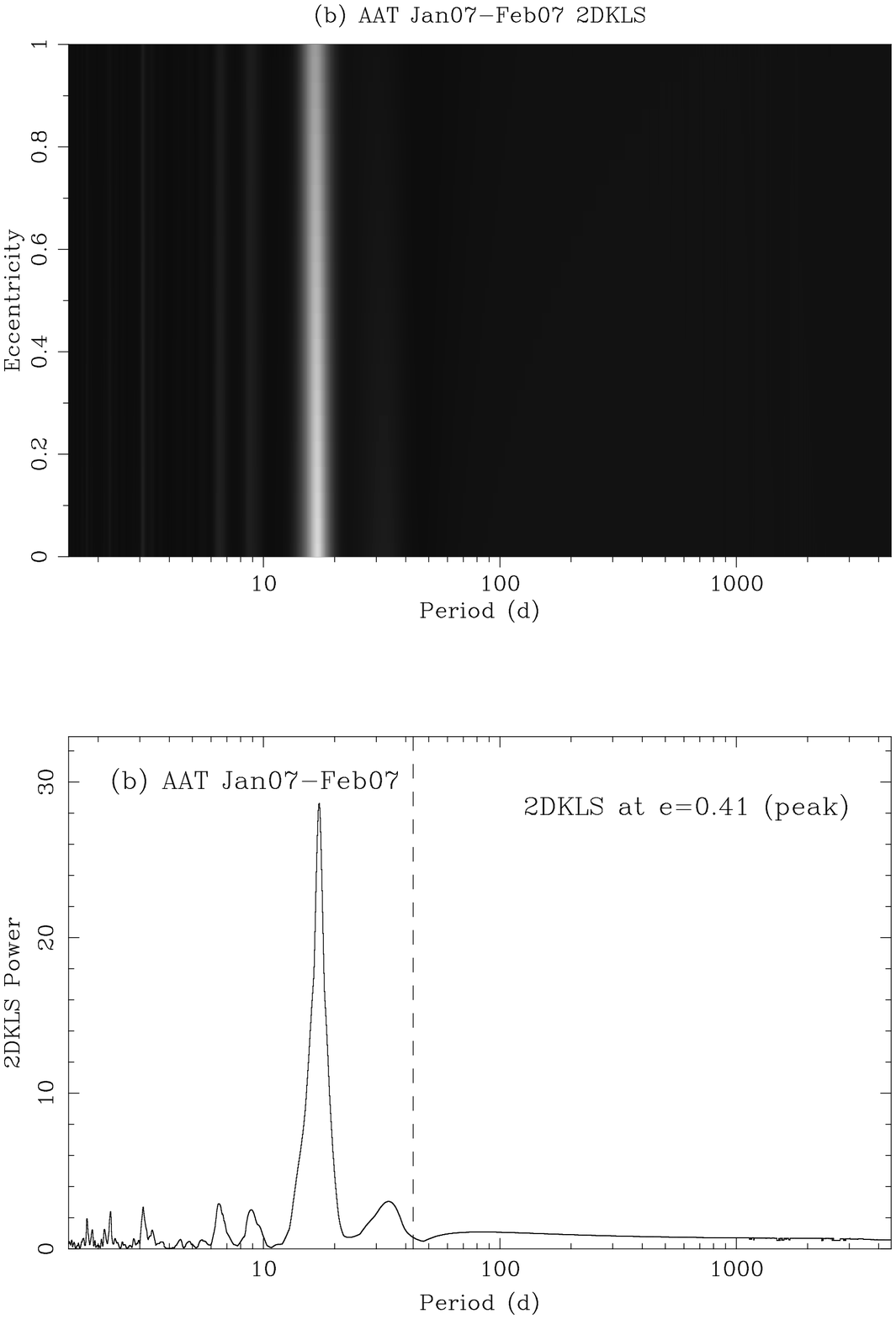}

\includegraphics[clip=true,width=8cm]{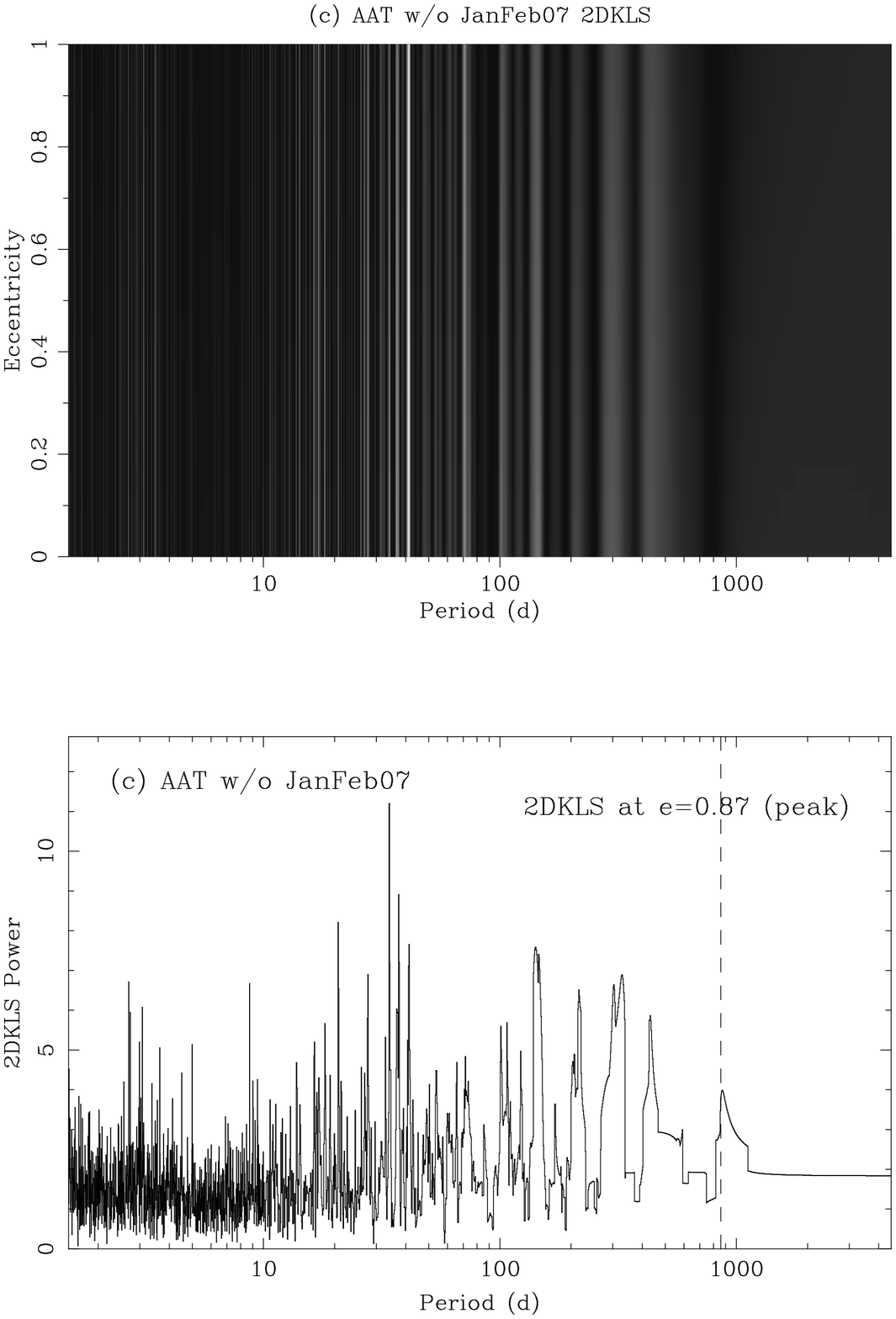}
\includegraphics[clip=true,width=8cm]{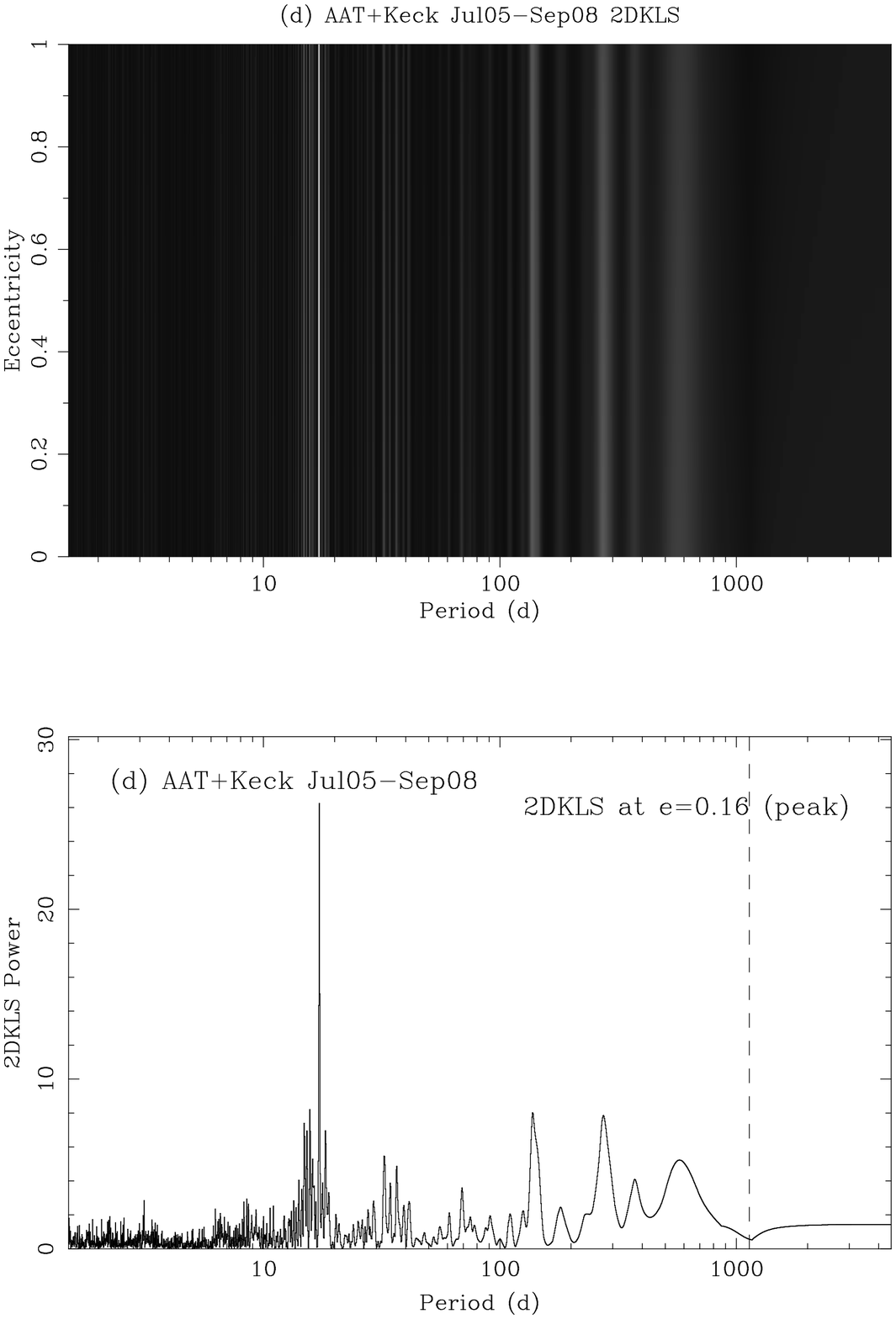}

\caption{Cuts through Two-Dimensional Keplerian Lomb-Scargle (2DKLS) periodograms 
for HD\,16417 at the eccentricities where the 2DKLS peaks, 
from velocities obtained (a) at the AAT 2005 Jul to 2007 Nov, (b) at the
AAT 2007 Jan-Feb (Rocky Planet Search), 
(c) at the AAT 2005 Jul to 2007 Nov, but {\em not} 
including velocities from 2007 Jan-Feb, and (d) at the AAT and Keck from 2005 Jul to 2008 Sep. 
The dashed vertical line in each panel is at the period corresponding to the length
of each data set.}
\label{2DKLS}
\end{figure}

The traditional Lomb-Scargle periodogram \citep{Lomb76,Scargle82}
estimates power as a function of period by fitting sinusoids to a data
set. The Two Dimensional Keplerian Lomb-Scargle periodogram 
\citep[2DKLS, ][]{AAPSXIV} extends this concept by fitting 
Keplerians as a function of both period and eccentricity.
We show in Figure \ref{2DKLS}, for three subsets of the HD16417 data,
slices through the 2DKLS at the eccentricity corresponding to
the peak power. The subsets examined are;
(a) all AAT data taken since 2005 Jul 27, 
(b) the AAT data taken in 2007 Jan and Feb, 
(c) all AAT data taken since 2005 Jul {\em except} that taken 
in 2007 Jan-Feb, and
(d) AAT and Keck data taken since 2005 Jul (with Keck data 
zero-point-corrected to the AAT system as described below). 
The first point to note is that
Fig. \ref{2DKLS} clearly shows evidence for
a periodicity at 17\,d. The second point to note is that that periodicity
is clearly seen in  just the 24 epochs obtained for HD\,16417 in the
major campaign run in 2007 (Fig. \ref{2DKLS}b). But perhaps most 
interestingly, if the long series of continuous data from 
that major run is removed (Fig. \ref{2DKLS}c), no convincing evidence
for any periodicity is detectable. The number of data points and the
internal measurement uncertainties of the data that produce 
Fig. \ref{2DKLS}b and \ref{2DKLS}c are almost exactly the same.
The difference is in the window function of the observations.
This is a key point to which we return below. And finally the Keck data 
(Fig. \ref{2DKLS}d) confirms and sharpens the 17\,d peak seen at the AAT.

\begin{figure}
\includegraphics[clip=true,width=12cm]{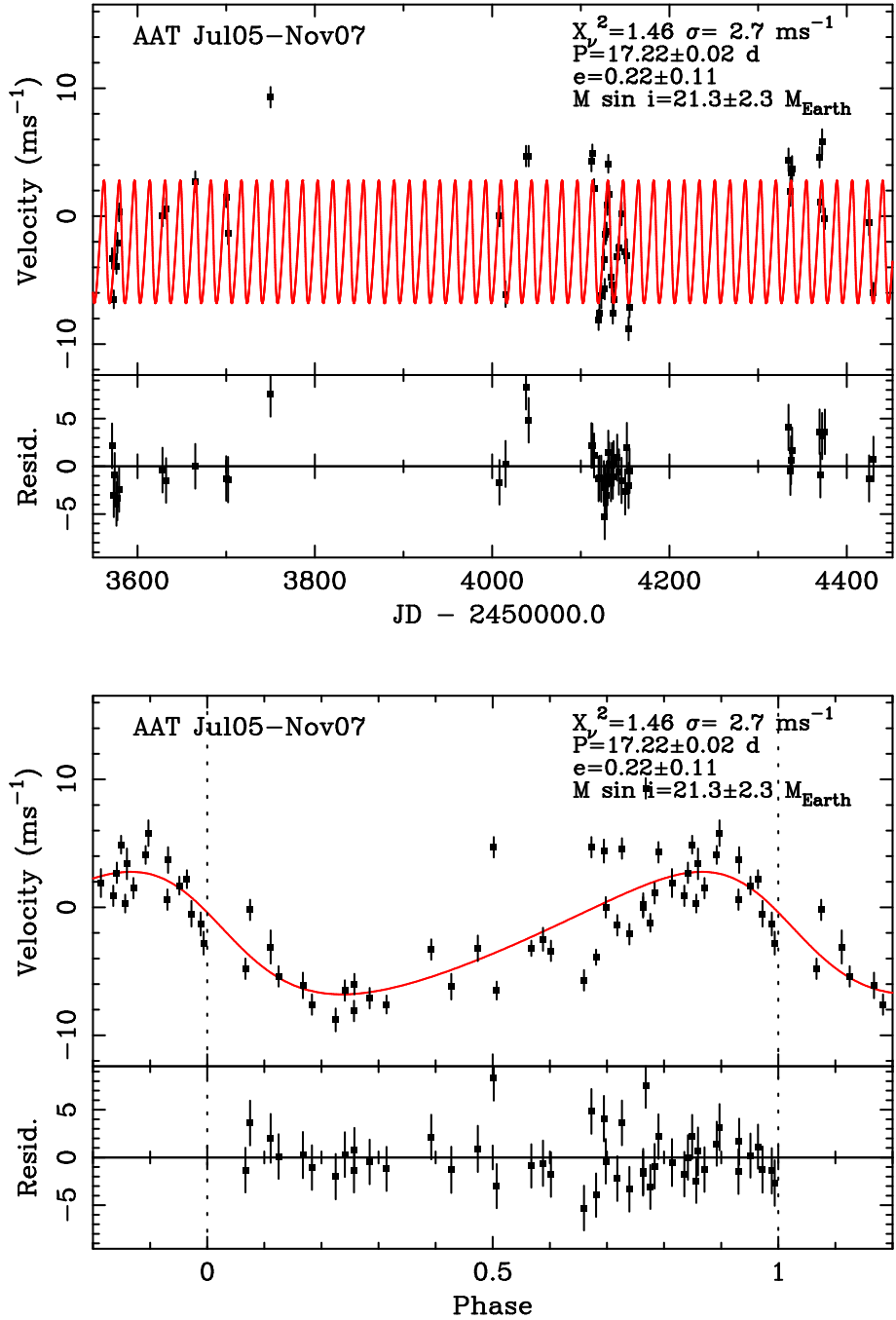}
\caption{Keplerian best fit to AAT data for HD\,16417 from 2005 Jul
to 2007 Nov, shown as both a function of time (upper plot) and phased at the
best-fit period (lower plot). The bars show the
internal measurement uncertainty produced by the Doppler measurement process.
In each plot the lower panel shows the residuals to the fit -- these bars
also include the jitter estimated for HD\,16417. A host star mass of 
1.2\msun\ and an intrinsic stellar
Doppler variability (i.e. jitter) of 2.2\ms\ are assumed.}
\label{fit}
\end{figure}

Using the 2DKLS to identify an initial period \citep{AAPSXIV}, 
a least-squares Keplerian fit 
to all AAT data obtained since 2005 Jul results
in the orbital parameters for HD\,16417b shown in Table \ref{orbit}.
Figure \ref{fit} displays this fit (and the residuals to it) as a function
of both time and orbital phase. The rms scatter to this fit is 2.7\,\ms, 
and the
reduced chi-squared ($\chi^2_\nu$) is 1.46. This fit indicates the 
presence of a planet with period 17.22$\pm$0.02\,d, eccentricity
0.22$\pm$0.11, semi-major axis 0.14$\pm$0.01\,AU and
minimum mass, \msini, 21.3$\pm$2.3\mearth. 

\begin{figure}
\includegraphics[clip=true,width=12cm]{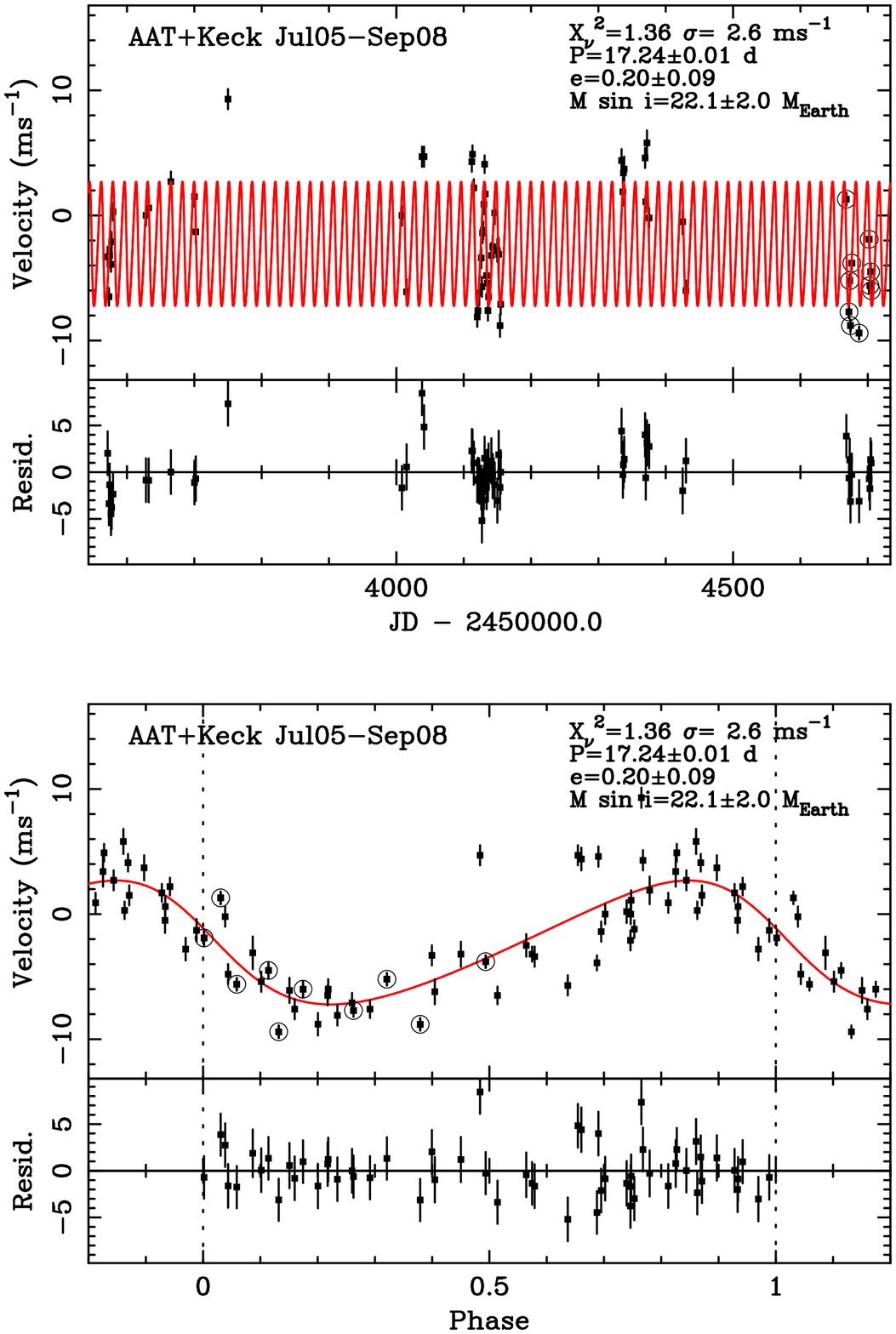}
\caption{AAT and Keck data (2005 Jul to 2008 Sep) phased at 
period obtained when a Keplerian is fit to both AAT and Keck data.
The Keck data are highlighted with circles. The bars shown  the
internal measurement uncertainty produced by the Doppler measurement process.
The lower panel shows the residuals to the fit -- these bars
also include the jitter estimated for HD\,16417. 
A host star mass of 1.2\msun\ and an intrinsic stellar
Doppler variability (i.e. jitter) of 2.2\ms\ are assumed.}
\label{bothfit}
\end{figure}

As an independent test of the validity of the Keplerian fit to the
AAT data, observations of HD\,16417 were acquired on 10 epochs in
2008 Aug-Sep with the HIRES spectrograph on the Keck~I
telescope. These data were processed as described by \citet{Vogt05}.
Being acquired with a completely different telescope,
spectrograph,  and detector system, these data provide an 
independent test of our AAT orbit.  
The Doppler observations from these 10 epochs have a different 
arbitrary velocity zero-point
from our AAT data, which we solve for by determining the mean difference
(5.30$\pm$0.8\,\ms) between them and the AAT Keplerian fit listed in Table \ref{orbit}. 
The Keck data have an rms scatter about the AAT Keplerian fit of 2.6\,\ms, and
are consistent with the AAT orbital fit.
The scatter of the Keck data about the AAT fit is consistent with the scatter seen about the AAT data, 
and is also consistent with being dominated by the 2.2\,\ms\ stellar 
jitter assumed for HD\,16417. A Keplerian fit to both the
AAT and Keck Doppler data is plotted in phased format in Fig. \ref{bothfit}
and has the parameters listed in Table \ref{orbit}. Inspection of the
Table shows that the AAT and AAT+Keck solutions are essentially identical.

\begin{figure}
\includegraphics[clip=true,width=12cm,angle=270]{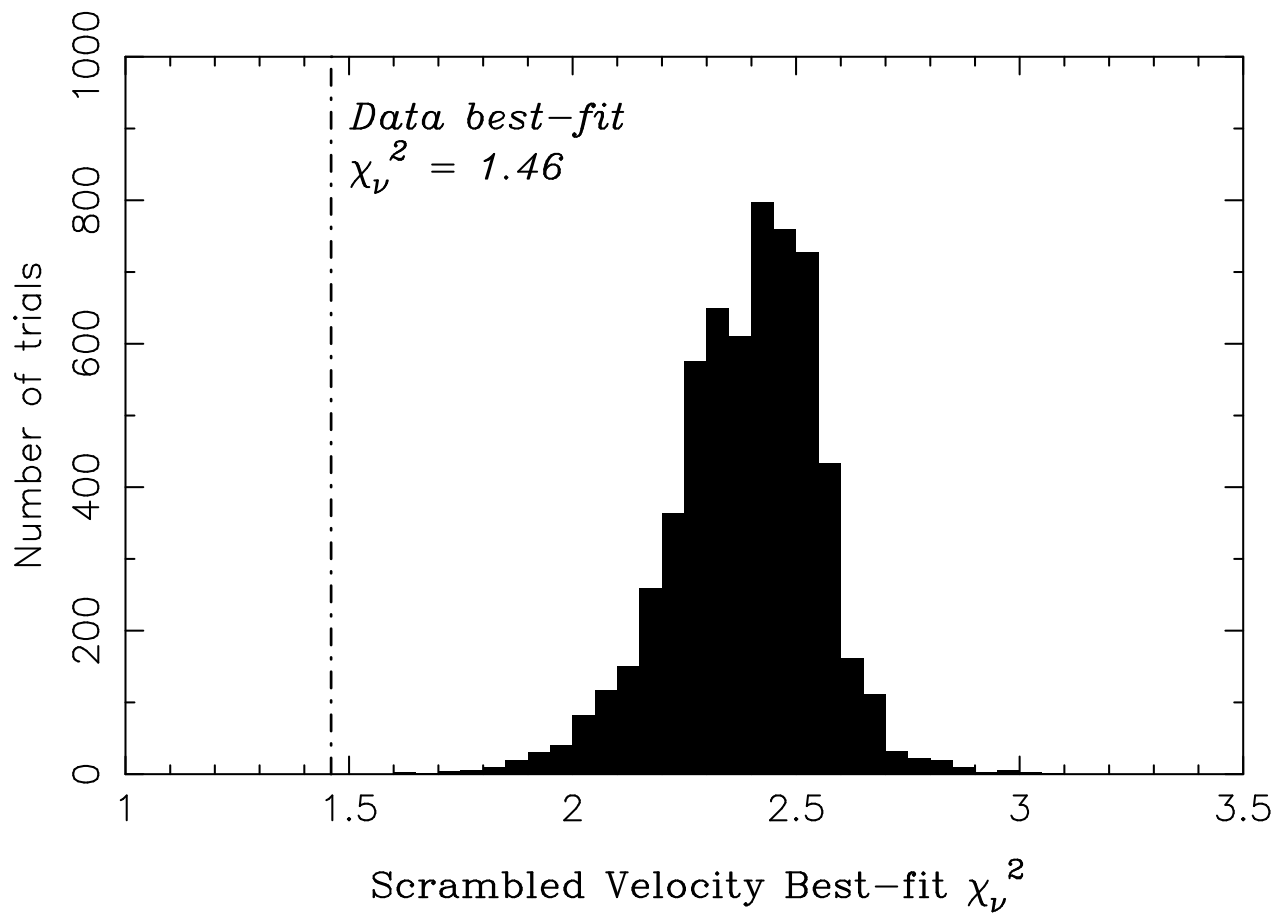}
\caption{Scrambled false alarm probability results. 
The histogram shows the $\chi^2_\nu$
values that result from the best Keplerian fits to 6000 
realizations of scrambled versions
of the AAPS velocities for HD\,16417. The dashed line 
shows the reduced $\chi^2_\nu$ for our actual data.}
\label{scrambled}
\end{figure}

To test the probability that the noise in our data set might have
resulted in a false  detection, we have run Monte Carlo simulations
using the ``scrambled velocity'' approach of \citet{marcy05}. This
technique makes the null hypothesis that no planet is present, and
then uses the actual data as the best available proxy for the combined
noise due to our observing system and the star. Multiple realizations
of that noise are then created by keeping the observed timestamps, and
scrambling the observed velocities amongst them. We created 6000
scrambled AAT velocity sets, and then subjected them to the same analysis
as our actual data set (i.e. identifying the strongest peak in the
2DKLS followed by a least-squares Keplerian fit).  No trial amongst 6000
showed a $\chi_{\nu}^2$ better than that obtained for the
original AAT data set, and the distribution of the scrambled reduced
$\chi_{\nu}^2$  (see Fig. \ref{scrambled}) shows a clear separation
from that obtained with the actual data. We conclude that the
probability of obtaining a false planetary detection from our
velocities of HD\,16417 is $<$0.017\%.

\section{Discussion}

The velocity semi-amplitude of HD\,16417b is quite low
($K=4.8$\,m\,s$^{-1}$), so we must consider the possibility that the
observed variation could be due to a stellar effect, such as a
rotating starspot, rather than a planet. Unfortunately, the velocity
amplitude is much too small for an analysis of line bisectors to
reveal any surface kinematics.  However, from the activity measure
log\,R$^{\prime}_{\rm HK}=-5.08$, we can predict a rotation period of
23-33\,d \citep{Wright04}. This is inconsistent with our measured
orbital period of 17.22\,d.   It is conceivable (as suggested by
\citet{Vogt05}  for the similarly short-period, low-mass planet
orbiting the inactive star HD\,190360) that a $\sim$17\,d Doppler
periodicity could be caused by {\em two} spot complexes at opposite
longitudes on a star with a rotation period of $\sim$34\,d.  However,
the presence of two such complexes would also wash out their Doppler
signal, such that each individual complex would need to be roughly
twice as large as that required to produce a similar velocity signal
from a single  complex. Given the implausibility of the contrived spot
features required on HD\,16417 to produce the observed Doppler
periodicity,  {\em and} the fact that the 17\,d periodicity has been
observed  to be coherent in phase over more than 3 years, we argue
that the  most probable explanation for the observed velocity signal
is a low-mass planet in a 17\,d orbit.

Given that multiple planet systems are being found around an increasing
number of extra-solar planet hosting stars \citep{butler06b}, 
we have carried out 
some simple tests of our data to see if further planets may be
present. The next most significant Doppler peak in our data (after the first
planet has been removed) is found at $\sim$290\,d.
Two-planet fits have been tested, and suggest the possibility of
a second highly-eccentric planet ($e > 0.8$) at P$\approx$289\,\d.
However, at present we are hesitant to propose this 
as a firm candidate given the low Doppler amplitudes involved.
The rms scatter about our single planet fit is just 2.6\,\ms\ (from
AAT and Keck data combined), which is consistent with being due to
our measurement uncertainties (1\,\ms) and jitter (2.2\,\ms) alone.
It is the nature of eccentric Keplerian fits that they are 
eminently capable of producing apparently good fits to roughly constant
data sets with a few velocity outliers -- however if those outliers
are truly due to noise, then such fits are essentially meaningless. 
As this is just the case we see here, 
more data will be required to confirm or deny the presence
of further planets in this system via the repeated observations
of periodic outliers to the single planet Keplerian solution.

The orbit of HD\,16417b appears to be non-circular ($e=0.20\pm0.09$), adding
to the growing list of short-period exoplanets with non-zero orbital
eccentricities. Tidal interaction with the planet host star is
expected to circularize the orbits of planets with short periods, with
circularization timescales typically shorter than the ages of their hosts.
We have used the relationship of \citet{GS66} to estimate the
circularisation timescale to be $\sim$350\,Gyr. This is much longer than
the upper limit to the age of HD\,16417 ($\sim$7\,Gyr). We used a
tidal quality factor, $Q_{\mathrm{p}}$, of $10^5$, which is in line
with the value estimated for solar system planets, and used a radius
estimate based on the measured radius of similar object HAT-P-11b
\citep{BTP09}.

The origin of the non-circular orbits is not entirely clear.
\citet{Matsumura08} suggested that either basing tidal
circularization calculations on our solar system is not appropriate,
or that these systems are affected by an external perturbation --
i.e. an outer (possibly undetected) planet. In the case of low-mass,
short-period exoplanets such as HD\,16417b, however, we advise
caution in placing too great an emphasis on non-zero
eccentricities. Two recent studies by \citet{otoole09} and
\citet{ST08}, have found that there is a bias \emph{against} measuring
zero-eccentricity orbits when signal-to-noise ratios are low. We
note that the fit uncertainty for HD\,16417b is also
quite high ($\sigma_e=0.1$), and so coupled with this bias, it is not
clear that the orbital eccentricity is well constrained. Monitoring of
the star is ongoing: this will provide future constraints
on all orbital parameters.

HD\,16417b raises the number of known planets with \msini\
minimum masses of Neptune-mass (or less) to eighteen. Interestingly,
roughly equal numbers have been found in orbit around G-, K- and M-dwarfs
(6, 6 and 6 respectively in  each spectral type). Interpreting these
numbers, though, is fraught with difficulty. On the one hand, low-mass
planets are easier to find around K- and M-dwarfs as their host star
masses are smaller. On the other hand, substantially fewer K- and
M-dwarf stars are under survey by Doppler programs at present, and
they tend to be fainter and more difficult to obtain optical Doppler
velocities for.

What we can clearly conclude from our observations of HD\,16417 is
that the efficiency of detecting low-mass planets in short period
orbits can be {\em significantly} enhanced through the use of
contiguous, targeted observing campaigns. As noted earlier, 24 epochs
of data on HD\,16417 obtained over a 48 night observing run show clear
evidence for the existence of a low-mass planet orbiting this star.
The same quality and quantity of data spread sparsely  over an 18
month period in observing blocks of 4-8 nights (and subject to the
exigencies of both telescope scheduling and weather) is {\em not} able
to detect the same planet. Such intensive observing -- extending
across dark, as well as bright, lunations -- may well need to become
the norm for future high-precision Doppler planet search observations
to continue probing to lower mass planets in short period orbits.

\acknowledgements

We acknowledge support from the following grants;  
NSF AST-9988087, 
NASA NAG5-12182, 
PPARC/STFC PP/C000552/1, 
ARC Discovery DP774000;
and travel support from the Carnegie Institution
of Washington and the Anglo-Australian Observatory.  
We are extremely grateful for the extraordinary support we 
have received from the
AAT technical staff  -- E. Penny, R. Paterson, D. Stafford,
F. Freeman, S. Lee, J. Pogson, S. James, J. Stevenson, K. Fiegert and G. Schaffer.  

\facility{AAT,Keck:I}

\onecolumn

\clearpage

\clearpage

\clearpage

\clearpage

\begin{deluxetable}{lrrrrrrc}
\tablenum{1}
\tablecaption{Properties of HD\,16417\label{hd16417_atm}}
\tabletypesize{\small}
\tablewidth{0pt}
\tablehead{
\colhead{Referece}      & \colhead{T$_{\rm eff}$} 
                                   & \colhead{[Fe/H]}   
                                             & \colhead{Mass}    & \colhead{log(g) }  & \colhead{Age}     & \colhead{v\,$\sin i$}
                                                                                                                      & \colhead{R$^\prime_{\rm HK}$}
}
\startdata
Gray et al. 2006        &  5745\,K & $+$0.00 &  \nodata           & 4.11              & \nodata           & \nodata   & -5.093 \\
Valenti \& Fischer 2005 &  5817\,K & $+$0.07 & 1.38$\pm$0.12\msun & 4.17              & 5.8$\pm$0.6\,Gyr  & 2.1\,\kms & \\
da Silva et al. 2006    &  5936\,K & $+$0.19 & 1.18$\pm$0.04\msun & 4.12$\pm$0.03     & 4.3$\pm$0.8\,Gyr  & \\
Bond et al. 2006        &  \nodata & $+$0.03 & \nodata            & 4.05              & \nodata           & \nodata   &  \\
Nordstr\"om et al. 2004 &  5649\,K & $-$0.01 & 1.10$\pm$0.04\msun & \nodata           & 7.6$\pm$0.7\,Gyr  & 2\,\kms   & \\
Jenkins et al. 2006     &  \nodata & \nodata & \nodata            & \nodata           & \nodata           & \nodata   & -5.08 \\
\enddata
\end{deluxetable}

\begin{deluxetable}{ccc ccc}
\tablenum{2}
\tablecaption{AAT \& Keck Velocities for HD\,16417\label{velhd16417}}

\tablewidth{0pt}
\tablehead{
\colhead{JD}             &  \colhead{RV}   & \colhead{Uncertainty} &\colhead{JD}             &  \colhead{RV}   & \colhead{Uncertainty}\\
\colhead{($-$2450000)}   &  \colhead{(\ms} & \colhead{(\ms)}       &\colhead{($-$2450000)}   &  \colhead{(\ms} & \colhead{(\ms)}      
}
\startdata
{\bf AAT data } &      &      &  4112.0116  &     4.3  &  0.8 \\
  831.0428  &    -4.9  &  1.3 &  4113.0269  &     4.9  &  0.7 \\
 1235.9478  &    -4.6  &  1.7 &  4115.0118  &     2.2  &  0.7 \\
 1383.3280  &    -6.2  &  1.3 &  4120.0503  &    -8.1  &  0.8 \\
 1527.0042  &    -6.2  &  1.5 &  4121.0305  &    -7.6  &  0.7 \\
 1745.2774  &    -3.0  &  2.1 &  4122.9829  &    -6.2  &  1.0 \\
 1918.9828  &     1.7  &  1.7 &  4125.9886  &    -3.4  &  0.8 \\
 2093.3381  &     2.5  &  1.3 &  4126.9851  &    -5.7  &  0.8 \\
 2152.1484  &    -6.4  &  1.7 &  4127.9908  &    -1.4  &  0.8 \\
 2187.2044  &   -13.0  &  1.6 &  4128.9865  &    -1.2  &  0.7 \\
 2510.2888  &    -6.2  &  1.7 &  4130.0112  &     0.9  &  0.8 \\
 2511.1718  &     4.5  &  1.8 &  4130.9868  &     4.1  &  0.7 \\
 2592.0257  &     3.4  &  1.7 &  4132.0035  &     1.7  &  0.7 \\
 2595.0301  &     4.0  &  1.8 &  4133.9993  &    -4.8  &  0.8 \\
 2653.9925  &    -7.8  &  1.5 &  4134.9980  &    -5.4  &  0.8 \\
 2654.9466  &    -5.3  &  1.2 &  4136.0016  &    -7.6  &  0.8 \\
 2709.9252  &    -4.2  &  1.6 &  4136.9966  &    -6.5  &  0.8 \\
 2858.3358  &   -12.8  &  2.3 &  4141.0055  &    -3.2  &  1.0 \\
 2859.2674  &    -9.0  &  1.4 &  4142.9713  &    -2.5  &  0.9 \\
 2943.1533  &     2.8  &  1.5 &  4145.9938  &     0.2  &  0.9 \\
 2947.1022  &    -0.3  &  1.5 &  4149.9572  &    -2.8  &  0.9 \\
 3008.0274  &    -3.4  &  1.4 &  4151.9790  &    -3.1  &  1.3 \\
 3042.9591  &    -1.8  &  1.4 &  4153.9422  &    -8.8  &  0.9 \\
 3044.9355  &     1.3  &  1.6 &  4154.9652  &    -7.1  &  0.8 \\
 3214.2841  &     0.6  &  1.6 &  4334.2403  &     4.4  &  0.9 \\
 3216.3303  &    -1.0  &  1.3 &  4336.2952  &     1.9  &  1.1 \\
 3243.3165  &    -7.5  &  1.5 &  4337.0800  &     3.4  &  1.2 \\
 3245.3017  &    -3.5  &  1.7 &  4338.3181  &     3.7  &  1.0 \\
 3281.1880  &    -4.1  &  1.4 &  4369.2268  &     4.6  &  0.8 \\
 3571.3031  &    -3.3  &  0.8 &  4370.2132  &     1.1  &  0.8 \\
 3573.2777  &    -6.5  &  0.7 &  4372.1686  &     5.8  &  1.0 \\
 3574.3199  &    -3.2  &  0.6 &  4375.2326  &    -0.2  &  0.8 \\
 3576.2722  &    -3.9  &  0.6 &  4425.1309  &    -0.5  &  1.0 \\
 3577.2801  &    -2.1  &  0.8 &  4430.0449  &    -6.0  &  0.8 \\
 3579.2940  &     0.3  &  0.7 &  {\bf Keck data} \\
 3628.2300  &     0.0  &  0.8 &  4668.1283  &     6.6  &  0.5 \\
 3632.2301  &     0.6  &  0.8 &  4672.1241  &    -2.4  &  0.5 \\
 3665.1501  &     2.7  &  0.8 &  4673.1315  &     0.1  &  0.5 \\
 3700.0923  &     1.5  &  0.8 &  4674.1344  &    -3.5  &  0.5 \\
 3702.1209  &    -1.3  &  0.9 &  4676.1059  &     1.5  &  0.5 \\
 3749.9836  &     9.3  &  0.8 &  4687.1134  &    -4.1  &  0.5 \\
 4008.2198  &    -0.0  &  0.8 &  4702.1000  &     3.4  &  0.6 \\
 4015.1835  &    -6.1  &  1.0 &  4703.0879  &    -0.3  &  0.5 \\
 4038.1589  &     4.7  &  0.8 &  4704.0405  &     0.8  &  0.6 \\
 4041.1044  &     4.7  &  0.8 &  4705.0833  &    -0.7  &  0.6 \\
\enddata
\end{deluxetable}

\begin{deluxetable}{lcc}
\tablenum{3}
\tablecaption{Orbital Solutions for HD\,16417b\label{orbit}}

\tablewidth{0pt}
\tablehead{
\colhead{Parameter} & \colhead{AAT\tablenotemark{a}} & \colhead{AAT+Keck\tablenotemark{b}}
}
\startdata
Orbital period $P$ (days)        & 17.22 $\pm$ 0.02   & 17.24 $\pm$ 0.01  \\
Velocity semiamplitude $K$ (\ms) &  4.8  $\pm$  0.5   &  5.0  $\pm$  0.4  \\
Eccentricity $e$                 & 0.22  $\pm$  0.11   & 0.20  $\pm$  0.09  \\
Periastron date (JD$-$2450000)   & 103.1 $\pm$  4.8   & 99.74 $\pm$  3.3  \\
$\omega$ (degrees)               &  70   $\pm$  29    &  77   $\pm$  26   \\
M$\sin i$ (\mearth)              & 21.3  $\pm$   2.3  & 22.1  $\pm$  2.0 \\
semi-major axis (AU)             &  0.14 $\pm$  0.01  &  0.14 $\pm$  0.01 \\ 
N$_{\rm fit}$                    &  50                &  60               \\
RMS (\ms)                        &  2.7               &  2.6              \\
\enddata   
\tablenotetext{a}{Solutions for all AAT data obtained since 2005 Jul 19}
\tablenotetext{b}{Solutions for all AAT data obtained since 2005 Jul
  19 and all Keck data}
\end{deluxetable}

\end{document}